\documentclass[aps,prl,twocolumn,superscriptaddress, showpacs]{revtex4}

\usepackage{graphicx}
\usepackage{times}

\begin{document}
\title{Magnetoresistance oscillations and relaxation effects at the SrTiO$_3$ | LaAlO$_3$ interface}

\author{M. van Zalk}
\thanks{M.v.Z. and J.H. contributed equally to this work.}
\affiliation{Faculty of Science and Technology and MESA$^+$ Institute for Nanotechnology, University of Twente, 7500 AE Enschede, The Netherlands}
\author{J. Huijben}

\affiliation{Faculty of Science and Technology and MESA$^+$ Institute for Nanotechnology, University of Twente, 7500 AE Enschede, The Netherlands}
\affiliation{Strategic Research Orientation NanoElectronics, MESA$^+$ Institute for Nanotechnology, University of Twente, 7500 AE Enschede, The Netherlands}

\author{A.J.M. Giesbers}
\affiliation{High Field Magnet Laboratory, Institute for Molecules and Materials, Radboud University Nijmegen, 6525 ED Nijmegen, The Netherlands}

\author{M. Huijben}
\affiliation{Faculty of Science and Technology and MESA$^+$ Institute for Nanotechnology, University of Twente, 7500 AE Enschede, The Netherlands}

\author{U. Zeitler}
\author{J.C. Maan}
\affiliation{High Field Magnet Laboratory, Institute for Molecules and Materials, Radboud University Nijmegen, 6525 ED Nijmegen, The Netherlands}

\author{W.G. van der Wiel}
\affiliation{Strategic Research Orientation NanoElectronics, MESA$^+$ Institute for Nanotechnology, University of Twente, 7500 AE Enschede, The Netherlands}

\author{G. Rijnders}

\author{D.H.A. Blank}

\author{H. Hilgenkamp}
\author{A. Brinkman}
\email[]{a.brinkman@utwente.nl}

\affiliation{Faculty of Science and Technology and MESA$^+$ Institute for Nanotechnology, University of Twente, 7500 AE Enschede, The Netherlands}

\date{\today}

\begin{abstract}
We present low-temperature and high-field magnetotransport data on SrTiO$_3$ | LaAlO$_3$ interfaces. The resistance shows hysteresis in magnetic field and a logarithmic relaxation as a function of time. Oscillations in the magnetoresistance are observed, showing a $\sqrt{B}$ periodicity, both in large-area unstructured samples as well as in a structured sample. An explanation in terms of a commensurability condition of edge states in a highly mobile two-dimensional electron gas between substrate step edges is suggested. 
\end{abstract}


\maketitle

The discovery of conducting interfaces between the insulating perovskites SrTiO$_3$ and LaAlO$_3$ \cite{ohtomo2004hme} has generated intensive research in recent years. The conduction arises from a charge redistribution, dubbed electronic reconstruction, that occurs at the interface in order to counteract an otherwise diverging electric potential in a polar material. So far, oxide interface samples exhibited metallicity \cite{ohtomo2004hme,huijben2006ecc,thiel2006tqt,basletic2007msd}, superconductivity below 200~mK \cite{reyren317sib} and magnetic hysteresis below 300~mK \cite{brinkman2007mei}. The large variety of transport properties can be understood by taking into account three structural aspects: the presence of oxygen vacancies \cite{siemons2007ocd, herranz2007hml, kalabukhov2007eov}, lattice deformations (including cation disorder) and the electronic interface reconstruction itself. The relative contributions are largely determined by the growth conditions \cite{reviewHuijben}.

Two-dimensional (2D) electron gases in semiconductor heterostructures have had an important impact on our fundamental understanding of electronic transport. Oxide 2D electron systems are particularly interesting because of the richness of electronic phases that the oxides provide. Two-dimensionality and the quantum Hall effect were recently demonstrated for ZnO/Mg$_x$Zn$_{1-x}$O heterostructures \cite{tsukazaki2007qhe}. For the SrTiO$_3$ | LaAlO$_3$ interface, support for the 2D character of the metallicity is provided by the substrate termination dependence of the conductivity \cite{ohtomo2004hme}, the interlayer spacing dependence of coupled interfaces \cite{huijben2006ecc}, the abrupt onset of conductivity above a critical LaAlO$_3$ thickness \cite{thiel2006tqt}, conducting atomic force microscopy across cleaved samples \cite{basletic2007msd} and the Kosterlitz-Thouless nature of the superconducting phase transition \cite{reyren317sib}. However, the quantum Hall effect has not yet been observed at 2D metallic SrTiO$_3$ | LaAlO$_3$ interfaces.

For LaAlO$_3$ deposited on SrTiO$_3$ at relatively high oxygen pressures (10$^{-3}$ mbar), electronic reconstruction is expected to be the dominant cause of the transport properties. In these samples, a low-temperature upturn in the interface resistance, magnetic hysteresis and a negative magnetoresistance have been observed \cite{brinkman2007mei}. Magnetism has been predicted theoretically \cite{okamoto2006lro,pentcheva2006clo}, but the shape of the observed hysteresis curve and the nature of the magnetic ordering could not yet be fully explained. Additionally, it is an intriguing open question how the 2D nature of the interface electron gas would interplay with the magnetic effects, and whether this could give rise to novel transport phenomena. 

\begin{figure}
\includegraphics[width=.8\columnwidth]{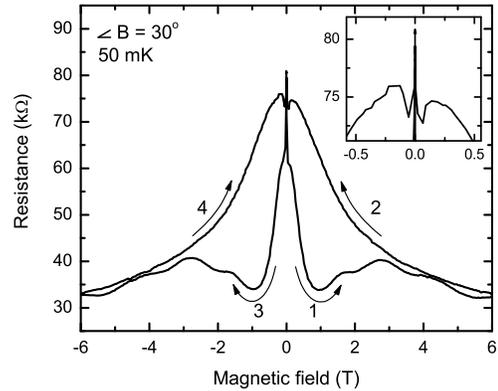}
\caption{\label{hysteresisloop} Magnetoresistance as a function of applied magnetic field at 50~mK, showing both hysteresis and oscillations. The numbered arrows denote the sweep direction. The inset shows the behavior around zero field. The sharp peak at zero field is observed both in the upsweep as in the downsweep. }
\end{figure}

In this Letter, we present transport measurements on SrTiO$_3$ | LaAlO$_3$ interfaces at 50~mK in magnetic fields, $B$, up to 30~T. The data provide insight in the nature of the magnetic phenomena and dimensionality of the transport. The hysteresis relaxes logarithmically as a function of time, suggestive of the presence of magnetic frustration. We observe magnetoresistance oscillations, which are periodic in $\sqrt{B}$, and not periodic in $1/B$, as is the case for the well-known Shubnikov-de Haas oscillations. A possible relation with the formation of edge states on substrate terrace edges is discussed. The presence of such states would imply the existence of a highly mobile 2D electron gas at the interface.


All samples discussed in this Letter are grown by pulsed laser deposition, as described in Ref.~\onlinecite{brinkman2007mei}. One of the samples is structured in order to improve uniformity over the probed surface. The structures are defined by Ar$^{+}$ etching, while keeping the etched surface insulating by optimizing the etch time and a short anneal in oxygen. Thin gold contact pads, defined by lift-off, are deposited to yield low contact resistances. The samples are electrically connected by wire bonding to the gold contact pads and, for the unstructured sample, to the edges of the sample.

Measurements were conducted in a dilution refrigerator with a base temperature of 40~mK in magnetic fields up to 30~T. Resistance measurements were performed using a standard low-frequency lock-in technique. Heating effects due to the measurement currents were excluded by using a measurement current of less than 5~nA, below which the resistance was current-independent.

\begin{figure}
\includegraphics[width=.8\columnwidth]{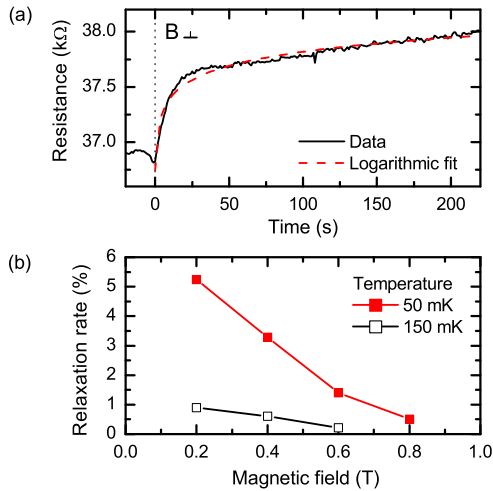}
\caption{\label{relaxation} (a) Time dependence of resistance after the magnetic field sweep is interrupted at 0.8~T (indicated by the vertical dotted line). The data are fitted by a logarithmic function (dashed line). (b) Relaxation rates obtained by fitting the time dependent resistance with a logarithmic fit. The lines are guides for the eye.}
\end{figure}

Figure \ref{hysteresisloop} shows a typical magnetoresistance curve at 50~mK. Clearly visible are the hysteresis and oscillatory behavior. At zero field, a sharp peak in the resistance is observed, which is suppressed rapidly as the field is increased. When the field is increased further, the resistance oscillates in field. In the downsweep, the resistance smoothly increases. Only when the background is substracted a small reminiscent of the oscillations can be found. At 0.1~T the resistance shows a minimum, followed by the sharp zero field peak. The hysteretic behavior is independent of the direction of the preceding field sweep, in contrast to ordinary ferromagnetic behavior. The overall magnetoresistance $[R(B)-R_0]/R_0$ increases with lower temperatures up to values of $-70$~\% ($-91$~\% for the structured sample). 

Within the hysteresis loop, an upward resistance relaxation is observed when the field is held constant. Figure~\ref{relaxation}a shows the resistance as a function of time, $t$, in a constant field of 0.8~T. Here, the field is perpendicular to the substrate, but the effect is independent of the field orientation. Relaxation rates on the order of seconds often indicate frustration and disorder, such as in spin glass states. The relaxation of these states can be described by an activation process with a wide distribution of energy barriers, yielding a $\ln{t}$ dependence of the resistance \cite{Moorjani_glasses}. A relaxation rate can be defined as $R_0^{-1} \frac{dR}{d\ln{t}}$, where $R_0$ is the resistance value at the start of the relaxation. The measured relaxation data can be fitted well with this logarithmic term and the extracted relaxation rates are presented in Fig.~\ref{relaxation} b. The relaxation rate decreases monotonically with temperature and field.

\begin{figure}
\includegraphics[width=.8\columnwidth]{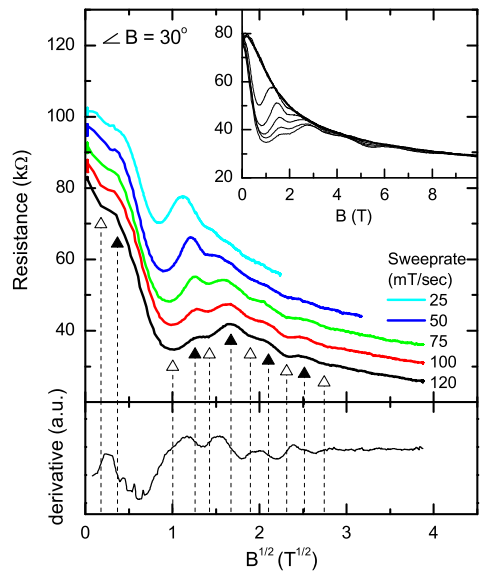}
\caption{\label{oscillations} Magnetoresistance oscillations for different sweep rates at 50~mK. In the main graph, the curves are offset by 5~k$\Omega$ per curve. Minima (maxima) positions are designated by open (closed) triangles and drop lines. The lower graph shows the derivative of the curve measured with a sweep rate of 120~mT/sec. The inset shows the complete series of field sweeps, plotted without offset.}
\end{figure}

The time-dependent resistance relaxation suggests a magnetic field sweep rate dependence of the hysteresis loop due to the balance between the relaxation rate and the field sweep rate. In Fig.~\ref{oscillations} magnetic field sweeps are shown for different sweep rates with the field oriented under an angle of 30$^\circ$ with respect to the sample surface. At higher sweep rates, the hysteresis loop opens further and resistance oscillations become visible. The minima and maxima positions shift only weakly with increasing sweep rate, excluding time-dependent oscillatory phenomena. Heating effects due to the high field sweep rates are unlikely, since for moderate sweep rates below 25~mT/sec small oscillations can already be observed. Furthermore, the resistance values in zero field and in high field do not vary in subsequent field sweeps. The amplitude of the oscillations is on the order of the resistance quantum, $h/e^2= 25.8$~k$\Omega$.  

\begin{figure}
\includegraphics[width=.8\columnwidth]{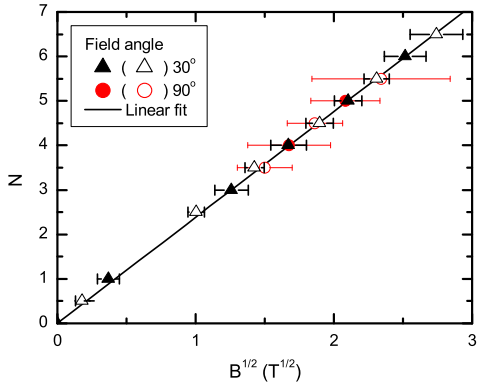}
\caption{\label{periodicity} Resistance minima and maxima versus $\sqrt{B}$. The minima and maxima positions show a square root behavior in magnetic field. The maxima (minima) positions are plotted at (half) integer values, indicated by filled (open) symbols. The circles correspond to the perpendicular orientation of the field, the angle between field and sample is 30$^\circ$ for the triangles.}
\end{figure}

The oscillatory behavior can be elucidated by plotting minima and maxima positions versus $\sqrt{B}$, see Fig.~\ref{periodicity}. A striking square root periodicity is found. Here, it is assumed that the second minimum and the third maximum in Fig. 3 are hidden in the steep resistance decrease around 0.7~T.

Although for perpendicular fields only low sweep rate data are available, we can extract resistance oscillations by subtracting a smooth background curve. For comparison, these data points are added to Fig.~\ref{periodicity}. It must be noted that the resistance of the sample has almost doubled between the two experiments, due to aging effects, so the directional dependent measurements have to be interpreted with great care. 

\begin{figure}
\includegraphics[width=.8\columnwidth]{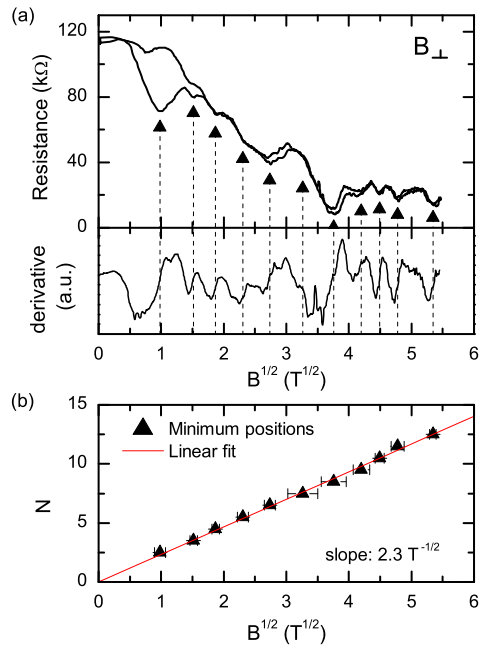}
\caption{\label{structured} (a) Magnetic field dependence of the resistance of a structured sample. Minima are denoted by triangles and dashed lines. (b) Also for structured samples, a square root dependence is found for the periodicity of the oscillations.}
\end{figure}

The oscillatory behavior with the $\sqrt{B}$ periodicity was reproduced on a 20~$\times$~20~$\mu$m$^2$ structured sample, see Fig.~\ref{structured}. Here, oscillations are observed up to the highest field of 30~T, both in increasing as well as in decreasing fields, even for sweep rates as low as 25~mT/sec. The amplitude of the hysteresis loop still scales with the field sweep rate. Subsequent sweeps reproduce well, although after a timescale of days changes occur in the relative amplitude of the oscillations accompanied by small shifts in the minima and maxima positions. 

Many oscillatory phenomena measured in magnetic fields are connected to the formation of Landau levels. The condition $\omega_c \tau \gg 1$, where $\omega_c=eB/m^*$ is the cyclotron frequency, $e$ the electron charge, $m^*$ the electron band mass and $\tau$ the elastic scattering time, cannot be fulfilled on first sight in our samples, given their low mobility \cite{brinkman2007sup}. However, pronounced resistance anisotropies are observed in our samples, which can be linked to the orientation of the substrate steps. In a sample with multiple structures, the anisotropy occurs in all structures individually, indicating that the anisotropy arises from the steps, rather than from an inhomogeneity over the sample. So far, the maximum resistance ratio measured for two perpendicular orientations implies that the resistance perpendicular to the steps is at least 10 times larger than the resistance in parallel to the steps. The step edge resistance might either arise from the discontinuity of the interface, acting as a scattering center or tunnel barrier, or from residual insulating \cite{ohtomo2004hme} SrO terminated regions between the TiO$_2$ terminated terraces, which are too narrow to be observed by atomic force microscopy.  If scattering occurs predominantly on the substrate step edges, a highly mobile electron gas between the steps is an intriguing possibility. 

In convential 2D electron gases, the formation of edge channels gives rise to zero longitudinal resistance and the quantum Hall effect. Whenever a Landau level crosses the Fermi energy, states connecting two adjacent edges become available and the zero resistance breaks down. This gives rise to $1/B$ periodic Shubnikov-de Haas oscillations. In principle, a confining potential modifies this picture, giving rise to a widening of the Landau level spacing in energy for low fields \cite{berggren1986mds}. However, for larger fields, one should recover the $1/B$ periodicity, which is not the case for our samples. The absence of $1/B$ oscillations might be explained by local variations in the Fermi energy. Within this picture, there will always be positions on the sample where a Landau level crosses the Fermi energy, and zero longitudinal resistances will no longer occur. However, edge channels of deeper lying Landau levels can still be formed. The radii of the skipping orbits are not determined by the Fermi energy, but by the field alone: $r_n = \hbar k_n/ eB = \sqrt{2(n+1/2)\hbar/eB}$, $k_n$ being the wave vector for Landau level $n$. The variations in carrier density will only cause small shifts of the skipping orbits towards or away from the edges. Within an edge channel the chemical potential is constant, and thus the formation of edge channels is expected to modify the potential distribution over the sample strongly. Although the exact physical character is unknown, the substrate steps likely provide some barrier, on which edge states can be formed. In our system, the probability for electrons to cross the edges and move into the next terrace is larger than zero. It is known that the barrier transparency can be greatly enhanced by the formation of edge channels \cite{muller1995qhe}. A scaling of the transparency, for example with the incident angle of the electron, seems not unlikely, giving rise to a commensurability condition $m\times2r_n = W$, where $m$ is some integer and $W$ the width of the terraces. The condition that an orbit should at least fit once between two adjacent edges is fulfilled for $m = 1$.

\begin{figure}
\includegraphics[width=.8\columnwidth]{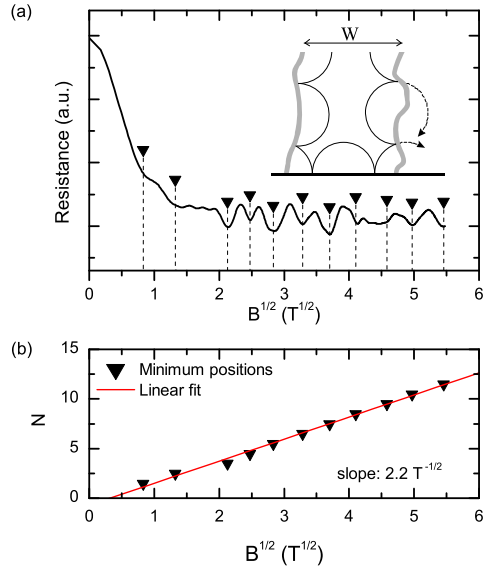}
\caption{\label{model} (a) Result of the edge-state model as described in the text. The inset shows a schematic representation of the edge states at the substrate step edges. Oscillations arise from the commensurability of the edge states between two adjacent semi-transparent step edges. (b) For a terrace width corresponding to that of the structured sample, the period of the oscillations and the $\sqrt{B}$ behavior is reproduced reasonably well.}
\end{figure}

We can qualitatively construct the expected oscillatory resistance behavior by making a summation of conductance peaks belonging to $n$ and $m$. Assuming that variations in the terrace width form the main broadening mechanism, the peak width can be obtained from $\Delta B = \frac{dB}{dW} \Delta W$. We add a minimum peak width arising from other broadening mechanisms. The influence of this for higher fields is negligible. The height of the peak is found by applying normalization. A typical result is given in Fig.~\ref{model} where a terrace width $W =$ 124~nm is used, which is the average terrace width of the structured sample as measured by atomic force microscopy and x-ray diffraction. The slopes $\Delta N/\Delta \sqrt{B}$ that can be determined this way correspond well to the experimentally obtained values for the different samples. 

In the scope of this interpretation, the fact that the first minimum in the unstructured sample can be seen from field values of 0.1~T sets a minimum to the mobility between the substrate steps of $10^5$~cm$^2$V$^{-1}$s$^{-1}$, given the condition $\omega_c \tau \gg 1$.

The sweep rate dependent measurements in Fig.~\ref{oscillations} clearly show a close connection between the hysteresis and the oscillations.  Upon increasing the temperature, the commensurability oscillations disappear at a rather low temperature. The oscillations have only been observed at 50~mK, whereas the hysteresis vanishes above 300~mK. The hysteresis can be explained by the ferromagnetic ordering of the spins of the conduction electrons, which can be expected for this type of interfaces \cite{okamoto2006lro,pentcheva2006clo}. The temperature dependence suggests that the ferromagnetic ordering is a prerequisite for the observation of oscillations. We also propose a periodical modulation of the magnetization itself (e.g. a homogeneous magnetization between the step edges and a disordered magnetization at the step edges) as a cause of commensurability oscillations. 

Although a possible interpretation of the unconventional periodicity in terms of step edge effects is suggested, we do not rule out the possibility for other causes which may include novel electron correlation effects. Interestingly, hysteresis and long-timescale relaxation effects are also observed in 2D quantum Hall ferromagnets \cite{eom2000qhf}.

In summary, the SrTiO$_3$ | LaAlO$_3$ interface shows intriguing oscillations in magnetic field. The strongly anisotropic resistance indicates an important contribution of the step edges to the resistance. A highly mobile 2D electron gas between the edges is now conceivable. A model has been suggested in which the observed $\sqrt{B}$ periodicity of the oscillations can be explained by the commensurability of edge states between step edges. Within the scope of this model, a lower limit for the mobility of $10^5$~cm$^2$V$^{-1}$s$^{-1}$ is estimated. When measurements can be performed on a single terrace, the quantum Hall effect might be observed for this system. The observation of a connection between the oscillations and the observed hysteresis and relaxation effects, suggests that magnetoresistance oscillations can only be observed in a ferromagnetically ordered state. The logarithmic relaxation of the resistance in the hysteresis loop indicates the presence of magnetic frustration in the system.

\begin{acknowledgments}
Enlightening discussions are acknowledged with T.H.~Geballe and A.F.~Morpurgo. This work is part of the research program of the Foundation for Fundamental Research on Matter (FOM), financially supported by the Netherlands Organization for Scientific Research (NWO), and the NanoNed program. 
\end{acknowledgments}

\bibliography{Manuscript}

\end{document}